\documentclass{aa}
\usepackage{natbib}
\bibpunct{(}{)}{;}{a}{}{,}
\usepackage{graphicx}
\usepackage[figuresright]{rotating}
\usepackage{txfonts}
\usepackage{longtable}
\begin{document}
\newcommand{\water}{H$_2$O}
\newcommand{\wlum}{$L_{{\rm H_2 O}}$}
\newcommand{\kms}{km s$^{-1}$}
\newcommand{\HII}{\hbox{H{\sc ii}}}
\newcommand{\HI}{\hbox{H{\sc i}}}
\newcommand{\solmass}{\hbox{M$_{\odot}$}}
\newcommand{\solum}{\hbox{L$_{\odot}$}} 
\defcitealias{newmas}{HPT}
    \title{Narrow-line Seyfert\,1 galaxies: an a{\bf $masing$} class of AGN
\thanks{Table 1 is only available in electronic form at http://www.aanda.org}}

 \author{A.\ Tarchi\inst{1}
         \and P.\ Castangia\inst{1}
         \and A. Columbano\inst{1,2}
         \and F.\ Panessa\inst{3}
         \and J.\ A.\ Braatz\inst{4}}

    \offprints{A.\ Tarchi}

 \institute{INAF-Osservatorio Astronomico di Cagliari, Loc. Poggio dei Pini, Strada 54, I-09012 Capoterra (CA), Italy\\
               \email{atarchi@oa-cagliari.inaf.it}
               \and Dipartimento di Fisica, Universit\'{a} degli studi di Cagliari, Cittadella Universitaria, 09042 Monserrato (CA), Italy\\
               \and IASF/INAF, via del Fosso del Cavaliere 100, 00133 Roma, Italy\\
               \and National Radio Astronomy Observatory, 520 Edgemont Road, Charlottesville, VA 22903, USA\\}
   
    \date{Received ; accepted }


  \abstract
  {Narrow-line Seyfert 1 (NLS1) galaxies are a class of active galactic nuclei (
AGN) that have all the properties of type 1 Seyfert galaxies but show peculiar characteristics, including the narrowest Balmer lines, strongest Fe II emission, and extreme properties in the X-rays. Line and continuum radio observations provide an optimal tool to access the (often) optically obscured innermost regions of AGN and reveal the kinematics of the gas around their central engines.}
  {We investigate the interplay between the peculiar NLS1 class of AGN and the maser phenomenon, to help us understand the nature of the maser emission in some NLS1s where water maser emission has been detected.}
  {We observed a sample of NLS1 galaxies with the Green Bank Telescope in a search for water maser emission at 22 GHz. We also reduced and analysed archival Green Bank Telescope and Very Large Array data and produced 22-GHz spectra for the five NLS1 galaxies with detected maser emission. In particular, we imaged the maser and nuclear radio continuum of NGC~5506 at subarcsec scales with the Very Large Array.}
  {We discovered maser emission in two NLS1 galaxies: IGR~J16385-2057, and IRAS~03450+0055. In addition to the three previously known maser detections in the NLS1s Mrk~766, NGC~4051, and NGC~5506, this yields a water maser detection rate in NLS1 galaxies of $\sim$ 7\% (5/71). This value rises significantly to $\sim$ 21\% (5/24) when considering only NLS1 galaxies at recessional velocities less than 10000 \kms. For NGC~4051 and NGC~5506, we find that the water maser emission is located within 5 and 12 pc, respectively, of nuclear radio continuum knots, which are interpreted as core-jet structures.}
 {The water maser detection rate in NLS1s is surprisingly high, much higher than the detection rate obtained for type 1 AGN and similar to those in Seyfert 2 and low-ionization nuclear emission-line region galaxies. The masers in NGC~4051 and NGC~5506 are nuclear and associated with the AGN, either with an accretion disk, a radio jet, or a nuclear outflow. The apparent lack of high-velocity maser features and evidence, recently reported, of radiative outflows and radio jets in the host galaxies seems to favour interpretation as a jet or an outflow. A similar association is also seemingly true for the maser in Mrk~766, IGR~J16385-2057, and IRAS~03450+0055, although, in these cases, without radio interferometric measurements we cannot rule out an off-nuclear origin of the emission.}


   \maketitle
%
  
\section{Introduction}

According to the widely accepted unified model of active galactic nuclei (AGN; e.g., \citealt{antonucci93}, \citealt{urry95}), in their very centres there is a supermassive black hole surrounded by a parsec-scale accretion disk. The disk emits intensely in the ultraviolet (UV) and soft X-ray wavelengths. A torus (or a thick disk) of atomic and molecular gas, with a size of 1-100 pc, surrounds the accretion disk and obscures the optical and UV emission along certain directions. Therefore, depending on the line of sight, the object appears as either a type 1 or type 2 AGN. In type 1 AGN, the accretion disk and black hole are viewed through the hole in the torus, while in type 2 AGN they are obscured by the torus.

Narrow-line Seyfert 1 (NLS1) galaxies have attracted particular attention. While Seyfert 2's are active galaxies with narrow emission-line optical spectra, NLS1s have the broad emission-line optical spectra of type 1 Seyfert galaxies, but with the narrowest Balmer lines from the broad line region and the strongest Fe II emission (e.g., \citealt{osterbrock89}, \citealt{veron01}). Extreme properties are also observed in X-rays, such as a strong soft excess emission below 1 keV and rapid flux variability, accompanied by steeper photon indices with respect to (w.r.t.) Seyfert 1 (Sy\,1) galaxies (e.g., \citealt{leighly99}, \citealt{gallo04}). The complex X-ray spectra often display signs of cold and ionized absorption, a partial covering of material along the line of sight, and/or (relativistic) reflection components from the inner side of the accretion disk (see \citealt{komossa08} for a review). Because of its extreme and ambiguous characteristics, this class of AGN seems to challenge the Unified Model, hence deserves a more detailed investigation. 

Fundamental studies of the central regions of AGN are complicated by the extremely small scales and complex structures of the nuclear components. Furthermore, in type 2 AGN in particular, the inner regions are often obscured at optical and UV wavelengths. Observations at infrared (IR, the band where most of the nuclear radiation absorbed by the torus is re-emitted), X-ray, and radio frequencies can, however, access these obscured regions. In particular, at radio wavelengths, water maser studies are a unique tool for investigating the structure and kinematics of the gas close to the black holes.

Most extragalactic \water\ masers are associated with AGN, where they have been related to three distinct phenomena. The first is that they may form directly in the nuclear accretion disk, where they can be used to trace the disk geometry and rotation velocities, and may reveal the enclosed nuclear mass (e.\ g.\ \citealt{kuo11}). In some cases, they are also being used to measure accurate distances to their host galaxies (for the most recent case of UGC~3789, see \citealt{braatz10}). Secondly, they are associated with radio jets, where they are either the result of an interaction between the jet(s) and a molecular cloud or a coincidental overlap along the line of sight between a warm, dense molecular cloud and the radio continuum of the jet (NGC~1052: e.\ g.\ \citealt{sawada08}; NGC~1068: e.\ g.\ \citealt{gallimore01}; Mrk~348: e.\ g.\ \citealt{peck03}), hence provide important information about the evolution of jets and their hotspots. Thirdly, they are associated with nuclear outflows, tracing the velocity and geometry of nuclear winds, as in the case of Circinus \citep{greenhill03}.

So far, more than 3000 galaxies have been searched for water maser emission and  detections have been obtained in about 150 of them (Braatz et al., in prep), the majority being Seyfert 2 (Sy\,2) galaxies. In this paper, we report the results of a new Green Bank Telescope (GBT) survey to search for water maser emission in objects belonging to the NLS1 class of galaxies and, by analysing the main characteristics and line features of the 5 NLS1 galaxies where water maser emission has been detected in the present and past surveys, we discuss the relation between the maser phenomenon and NLS1 galaxies in the framework of the unified scheme of AGN.


\section{Observations and data reduction}\label{obs}

Between Feb. 5 and Feb. 12, 2011, we used the GBT\footnote{ The National Radio Astronomy Observatory is a facility of the National Science Foundation operated under cooperative agreement by Associated Universities, Inc.} to observe a sample of 52 NLS1 galaxies taken from the list of \citealt{veron01} in a search for 22 GHz water maser emission. We used the dual beam 18--22 GHz receiver in a total power nod mode, keeping one of the two beams on source during integration. The spectrometer was configured with two 200 MHz IFs. The first spectral window was centred on the systemic recession velocity of the target galaxy and the second was offset by 180 MHz to the red, giving a total coverage of 380 MHz (5100\,km\,s$^{-1}$). In our analysis we also incorporate archival GBT observations of masers in the NLS1 galaxies Mrk~766 (NGC~4253), NGC~4051, NGC~5506, and IGR~J16385-2057 (hereafter IGR~J16385). These observations were made with the same setup that we used for our survey.

We retrieved archival Very Large Array (VLA)$^{1}$ observations of the water maser line for both NGC~4051 and NGC~5506. For NGC~4051, A-array configuration data were available (project \#AH812, observed in July 2007), while for NGC~5506 the observations were performed in both DnC and A configurations (project \#AB741, observed in Feb. 1995, and project \#AB859, observed in Apr. 1998, respectively). We also incorporate broad-band VLA A-array 22-GHz continuum data of NGC~5506 (project \#AC524, observed in Jun. 2003). In addition, again from the NRAO archive, we retrieved A-array X-band continuum data for Mrk~766 (project \#AM384, observed in Jan. 1993).  

The GBT and VLA data were reduced and analysed using the Green Bank Telescope Interactive Data Language (GBTIDL; for a detailed description on the procedures adopted, see \citealt{braatz10}) and Astronomical Image Processing System (AIPS) packages, respectively. We generated images of the VLA water maser spectra of NGC~4051 and NGC~5506 using the Continuum and Line Analysis Single-dish Software (CLASS), a package implemented in the Grenoble Image and Line Data Analysis Software (GILDAS).

%

\section{Results}\label{res}
  
\begin{figure}
\centering
\includegraphics[width= 9.2 cm]{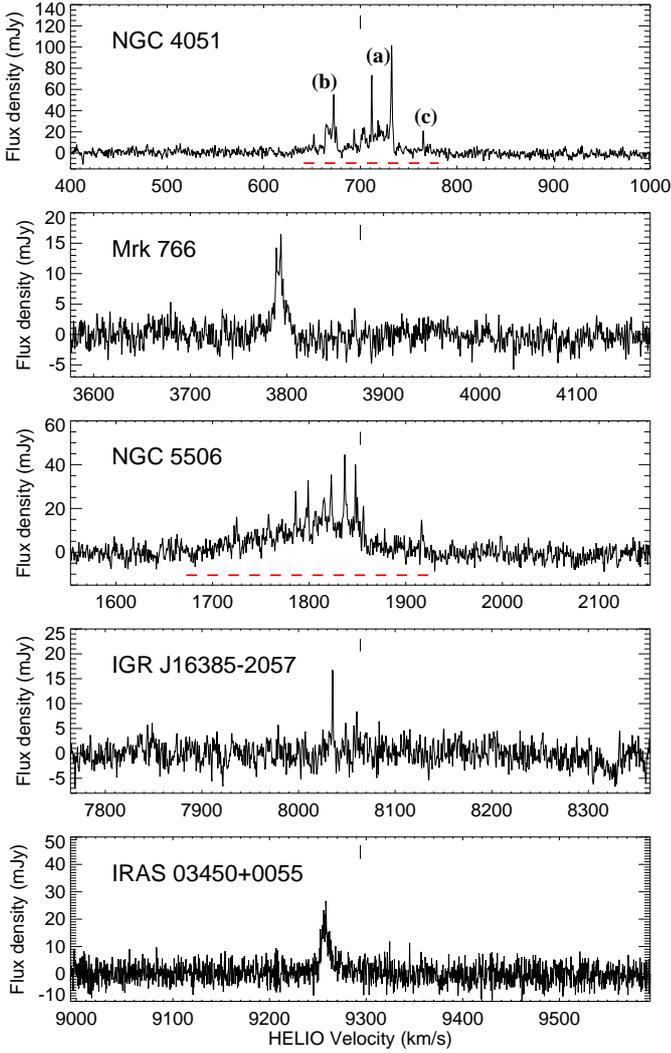}
\caption{GBT spectra of the five NLS1 galaxies where \water\ maser emission has been detected so far. The vertical line segment indicates the recessional velocity of the galaxy. The red dotted lines in the spectra of NGC~4051 and NGC~5506 indicate the velocity ranges covered by the VLA spectra. The (a), (b), and (c) labels for NGC~4051 indicate the three groups of features discussed in Sect.~\ref{origin}.}
\label{fig:threespectra}
\end{figure}

In Table~\ref{all}, we report the results of our GBT search for water masers in the 52 NLS1 galaxies taken from \citet{veron01}. From this sample, we detected one galaxy, IRAS~03450+0055 (hereafter IRAS~03450; Fig. 1). We also present GBT maser spectra in four other NLS1 galaxies (Fig.~\ref{fig:threespectra}). For NGC~4051 and NGC~5506, the masers were initially reported by \citet{hagi03} and \citet{braatz94}, respectively. The maser in Mrk~766 was discovered by the Megamaser Cosmology project\footnote{ https://safe.nrao.edu/wiki/bin/view/Main/MegamaserCosmologyProject} and the maser in IGR~J16385 was discovered as part of our companion survey.

The single-dish spectra of all five detections display little or no evidence of high-velocity lines (Fig.~1). The spectra of NGC~4051 and NGC~5506 show multiple components, clustered near the systemic velocity. Multiple maser components may also be present in IGR~J16385 just above the detection limit. The spectra of Mrk~766 and IRAS~03450 contain instead a single, relatively broad feature blue-shifted w.r.t. the systemic velocity.

The main parameters of the maser lines, together with some relevant characteristics of the five host galaxies, are summarized in Table~\ref{lines}.

\begin{figure}
\centering
\includegraphics[width= 9 cm]{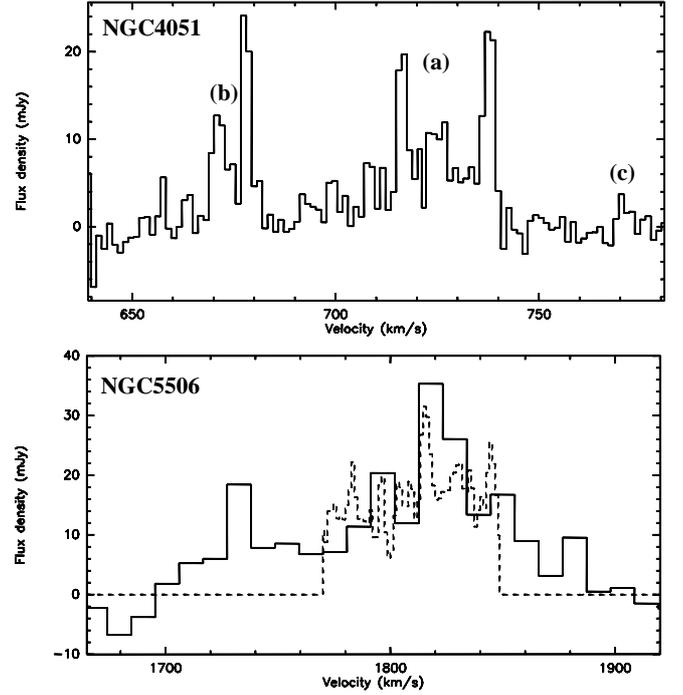}
\caption{{\it Upper panel}: the water maser spectrum of NGC~4051 produced by the combination the two separate IFs used in the VLA A-array observation. The letters indicate the main features constituting the line complex. The (a), (b), and (c) labels in the spectrum indicate the three groups of features discussed in Sect.~\ref{origin}. {\it Lower panel}: VLA spectra of the maser emission in NGC~5506, taken with the VLA DnC (solid line) and A (dashed line) arrays, respectively. Given the narrow band used in the VLA A-array measurements, no line-free channels were present. Hence, the spectrum shown is most likely the portion of maser emission in that velocity range plus a small amount of continuum emission. The zero-flux density line has been added to help visualize this aspect.}
\label{fig:twospectra}
\end{figure}

Figure~\ref{fig:twospectra} (upper panel) displays the \water\ maser spectrum of NGC~4051 observed with the VLA A-array. The spectrum is consistent with that of \citet{hagi07} and consists of several features  spanning a velocity range of $\sim$ 120 \kms. The emission arises from an unresolved spot with coordinates RA$_{J2000}$ = 12h03m09\fs61, Dec.$_{J2000}$ = +44d31'52\farcs7. The absolute position uncertainty is 0.1", dominated by the position accuracy of the phase calibrator (e.g. see \citealt{tarchi11} and references therein).

Figure~\ref{fig:twospectra} (lower panel) shows the spectra of the water maser emission in NGC~5506 observed with the VLA DnC (solid line) and A (dashed line) arrays. With the DnC observations, we detected \water\ maser emission from a number of features with velocities and flux densities consistent with those of the single dish. All of the emission arises from the position RA$_{\rm J2000}$ = 14$^{\rm h}$ 13$^{\rm m}$ 14\fs86, Dec$_{\rm J2000}$ = --03$^{\rm \circ}$ 12$^{\rm \prime}$ 27\farcs9, with a positional uncertainty of 0\farcs3. Owing to the relatively coarse resolution, the uncertainty in this case is dominated by the statistical errors that can be estimated by the synthesized beam size divided by the signal-to-noise ratio (for details, see \citealt{hagi01}). The location of the water maser emission derived from the A-array measurements is consistent with that of the DnC array (RA$_{\rm J2000}$ = 14$^{\rm h}$ 13$^{\rm m}$ 14\fs87, Dec$_{\rm J2000}$ = --03$^{\rm \circ}$ 12$^{\rm \prime}$ 27\farcs8. In this case, the observation was made with the VLA A-array and the position accuracy is 0\farcs1, determined by the uncertainty in the position of the phase calibrator. Unfortunately, the velocity coverage of the A-array measurement was too narrow to cover all of the maser lines. In addition, no line-free channels were present in the spectrum, making it impossible for us to properly subtract the continuum and differentiate the contribution of the continuum emission from that of the line. However, from the similarities with the same portion of spectrum in the DnC data (in particular the feature at V $\approx$ 1820 \kms; Fig.~\ref{fig:twospectra}), we are confident that the detected emission throughout the band is dominated by the water maser in NGC~5506 and its location is therefore reliable.

%

\section{Discussion}\label{disc}

\subsection{The water maser detection rate in NLS1 galaxies}
\label{detection}

Identifying water maser emission associated with Sy\,2 galaxies is not particularly surprising, given that the unified scheme for AGN requires an obscuring structure, which is probably an edge-on disk or torus, along the line of sight towards the nucleus of a type 2 AGN. This structure can provide a molecular reservoir and the amplification paths necessary for maser action. The relation between a type 1 Seyfert and the maser phenomenon is less obvious. According to the same paradigm, in type 1 objects, accretion disks and/or tori, if present at all, should be orientated face-on, making them less likely to produce detectable maser emission. Accordingly, out of the 150 water maser sources detected so far, maser emission has been detected in only very few type 1 Seyfert galaxies. Interestingly, two galaxies among these are classified as NLS1s: NGC~4051 and NGC~5506. In view of the above considerations, the association between water maser emission and the NLS1 classification is clearly intriguing. We searched for water maser emission with the GBT in a sample of 17 Seyfert galaxies (classified by \citealt{masetti08}) discovered by INTErnational Gamma-Ray Astrophysics Laboratory (INTEGRAL) as sources of hard X-rays between 20 and 40 keV \citep{bird10}. We detected water maser emission in one object (Fig.~1), IGR~J16385. Interestingly, this was the only NLS1 in the sample \citep{masetti08}. Further details on a statistical study of this campaign will be presented elsewhere (Castangia et al. in prep).

Galaxy classifications are taken from \citet{veron10}, with the exception of NGC~5506, which was (re)classified by \citet{nagar02} using near-IR spectroscopy. Nagar et al. detected the permitted OI $\lambda$1.1287 $\mu$m line (with full width at half maximum $<$ 2000 \kms) and the ``1 micron Fe II'' lines to identify the galaxy as a NLS1. Their finding raises the important issue that other galaxies may be misclassified, depending on the quality and type of spectral information available. In principle, according to \citet{nagar02}, other X-ray bright and highly variable ``type 2'' Seyferts, such as NGC~5506, may actually be partially obscured NLS1s. This is just a cautionary note that needs to be kept in mind when dealing with studies similar to the one proposed here.

Including all past surveys (Braatz et al.\ in prep) and our new detections in IGR~J16385 and IRAS~03450, a total of 71 NLS1 have been searched for water maser lines with five successful detections (Table~\ref{all}). This yields a detection rate of $\sim$ 7\%, which is comparable to the global rates of AGN surveys (e.\ g.\ \citealt{braatz97}). While this result is surprising, the maser detection rate in NLS1s becomes even more impressive when we consider a volume-limited sample that somewhat minimizes the limitation in sensitivity of the survey(s). When considering only NLS1 galaxies at recessional velocities less than 10000 \kms, the detection rate goes up dramatically to $\sim$ 21 \% (5/24). This value approaches the highest detection rates ever obtained for similarly volume-limited samples, in any class of AGN. For example, the most prolific class of AGN, the Sy\,2 galaxies, have detection rates of up to $\sim$ 20\%, while in `pure' Sy\,1 galaxies the percentage of water maser detections never exceeds 1\% (e.\ g.\ \citealt{braatz04};\cite{bennert09}; \cite{zhang10}; \citealt{ramolla11}).

\subsection{The origin of water maser emission in NLS1 galaxies}
\label{origin}

\begin{figure*}
\centering
\includegraphics[width= 17 cm]{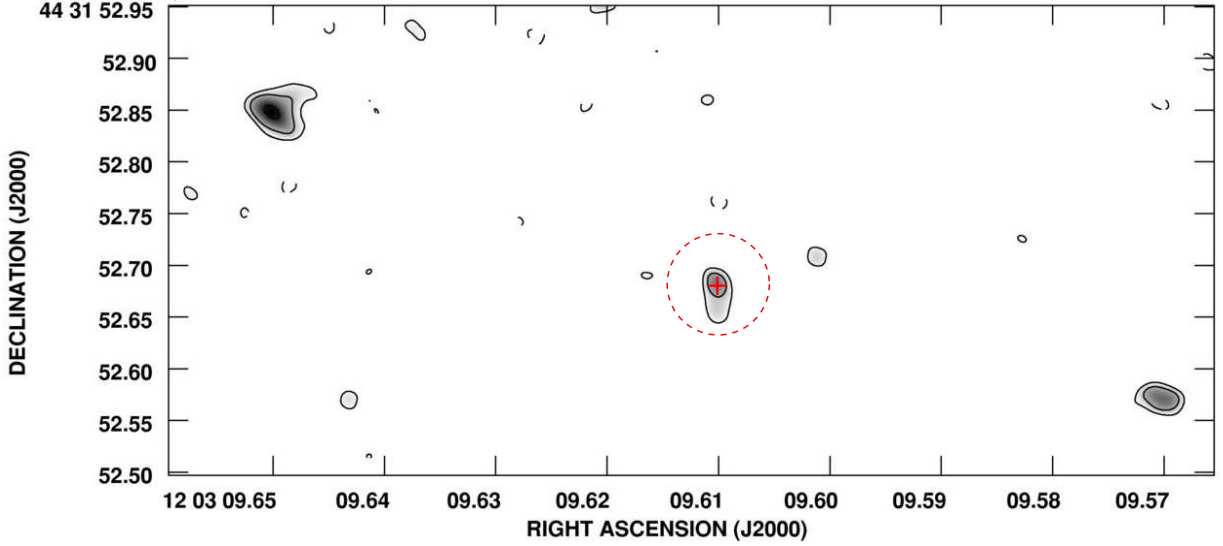}
\caption{1.6-GHz EVN radio continuum maps of the nuclear region of the NLS1 NGC~4051 (\citealt{giroletti09}). The size of the beam is 13.5 $\times$ 11.3 mas. The cross indicates the position (RA$_{J2000}$ = 12h03m09\fs61; Dec.$_{J2000}$ = +44d31'52\farcs7) of the maser derived with the VLA A-array (\cite{hagi07}; present work). The circle represents the 0\farcs1 arcsec VLA nominal position accuracy of the maser emission.}
\label{fig:n4051}
\end{figure*}

Water masers associated with an AGN accretion disk exhibit a typical spectrum with three distinct groups of features. One ``cluster'' of lines is found at the recessional velocity of the galaxy, while the other two are blueshifted and redshifted by hundreds of \kms (e.g., for NGC~4258: \citealt{miyoshi95}; for UGC~3789: \citealt{braatz10}). Water masers associated with radio jets (like in M~51, NGC~1068, Mrk~348, and NGC~1052) instead show single, broad (with linewidths of $\sim$ 100 km/s) maser lines offset from the systemic velocity of the host galaxies (e.g., see \citealt{lo05} and references therein). In the Circinus galaxy, the only object where some of the maser emission has been confidently associated with a nuclear outflow, the maser features associated with the outflow have relatively narrower widths (a few \kms), while line velocities are again found to be both red- and blue-shifted w.r.t the systemic velocity (\citealt{greenhill03}). 

The characteristics of the maser lines found in NLS1 galaxies (Sect.~\ref{res}) and, in particular, the lack of clear evidence for high-velocity lines (Fig.~\ref{fig:threespectra}) suggests that masers in NLS1s differ from the disk masers in classic Sy\,2, perhaps being associated with jets and/or nuclear outflows. However, it has to be kept in mind that the maser lines associated with jets, especially in Mrk~348 and NGC~1052, are typically (much) broader than those in NLS1s (100 vs. 10 \kms). In addition, the displacements of the line velocities from the systemic velocity in jet-maser galaxies are larger than those in NLS1s (120-150 vs 40-80 \kms) and all towards the red, with the only exception of NGC~1068 (\citealt{gallimore01}). Detailed studies of each case are thus necessary to unveil the origin of the maser emission in the five NLS1 targets of this work.

There are presently only two maser NLS1 galaxies with available interferometric measurements of both the radio continuum (with Very Long Baseline Interferometry; VLBI) and the maser emission (with the VLA): NGC~4051 and NGC~5506. For both galaxies, the nuclear radio continuum emission is resolved into compact knots that can be interpreted as core-jet structures (\citealt{giroletti09}, for NGC~4051; \citealt{middelberg04}, for NGC~5506). 

In the following, we consider, in detail for these two galaxies, the association of the maser emission with either accretion disks, jets, or outflows and we investigate the main characteristics of the five maser spectra found in NLS1s and their host galaxies to more tightly constrain the nature of the maser emission. The investigation of these five cases constitute a first attempt to understand the possible relation between NLS1 objects and the maser phenomenon.

\subsubsection{NGC~4051}
The nuclear radio source detected at both 1.6 and 5 GHz with the European VLBI Network (EVN) by \citet{giroletti09} is most easily described in terms of a jet-base/outflow phenomenon. The detection by Giroletti \& Panessa of three aligned sub-millijansky components in NGC 4051 suggests an ejection process. In particular, the position of the EVN radio core is coincident with that derived for the \water\ maser by \citet{hagi07} and ourselves, within 0\farcs1 (Sect.~\ref{res}), corresponding to 5 pc at the distance of NGC~4051 (Fig.~\ref{fig:n4051}). This suggests that the water maser emission arises from a molecular disk or from a nuclear jet/wind associated with the brightest radio continuum knot. This latter option seems to be consistent with the existence of a weak, relatively extended ($\sim$ 60 pc) jet in NGC~4051 (\citealt{jones11}, \citealt{king11}, and \citealt{maitra11}) and a shocked outflow (\citealt{pounds11} and \citealt{vaughan11}). In addition, the single-dish maser spectrum of NGC~4051 is somewhat reminiscent of that in Circinus where some of the maser emission is associated with a nuclear outflow \citep{greenhill03}.  

{\it {An alternative na\"{i}ve interpretation:}} While the jet/outflow association of the maser emission is indeed the most likely option, a speculative interpretation of the maser spectrum shown in Fig.~\ref{fig:threespectra} can be made in terms of an accretion disk maser. The maser spectrum could be described by three main peaks at velocities of $\sim$ 670, 720, and 770 \kms (see Fig.~\ref{fig:threespectra}) roughly centred on the systemic velocity. The system thus resembles the triple peak profile of accretion disk masers, albeit with much lower velocity offsets. The 'triple-peak profile' is also somewhat confirmed by the VLA A-array spectrum taken at a different epoch (Fig.~\ref{fig:twospectra}). In this scenario, we would obtain a rotation velocity of $V_{\rm rot} = V_{\rm obs}\,\cdot\,(\sin^{-1} i)$ $\sim$ 50$\,\cdot\,(\sin^{-1} i)$ \kms. From \cite{uttley04}, we obtain $M_{\rm BH}$ = 5 $\times$ 10$^{5}$\,M$_{\odot}$ as the mass  of the nuclear engine (Table~\ref{lines}). Combining these two parameters and assuming Keplerian rotation (as in NGC\,4258), this yields  
 an angular size of 18\,$\cdot$\,($\sin^{-2} i$) mas (see \citealt{tarchi07a} for details on a similar calculation), corresponding to a galactocentric radius of $R_{\rm GC}$ $\sim$ 0.9$\,\cdot\,(\sin^{2} i)$ pc. Assuming a nuclear accretion disk seen approximately edge-on ($i$ $\sim$ 90$\degr$), the radius of the accretion disk in NGC~4051 would be almost ten times larger than the one found, for example, in NGC\,4258 (0.1 pc). Such a large accretion disk in a Seyfert galaxy has, so far, only been directly measured in the Seyfert 1.9 galaxy NGC~1194, which has an inner and outer radius for the disk of 0.54 and 1.3 pc, respectively \citep{kuo11}. Alternatively, a possibly lower value for the inclination of the disk and/or for the black hole mass in NGC~4051 would result in a size of the disk more consistent with the one in NGC~4258. Follow-up VLBI observations have been requested aimed at shedding light on the aforementioned scenario.

\begin{figure*}
\centering
\includegraphics[width= 17 cm]{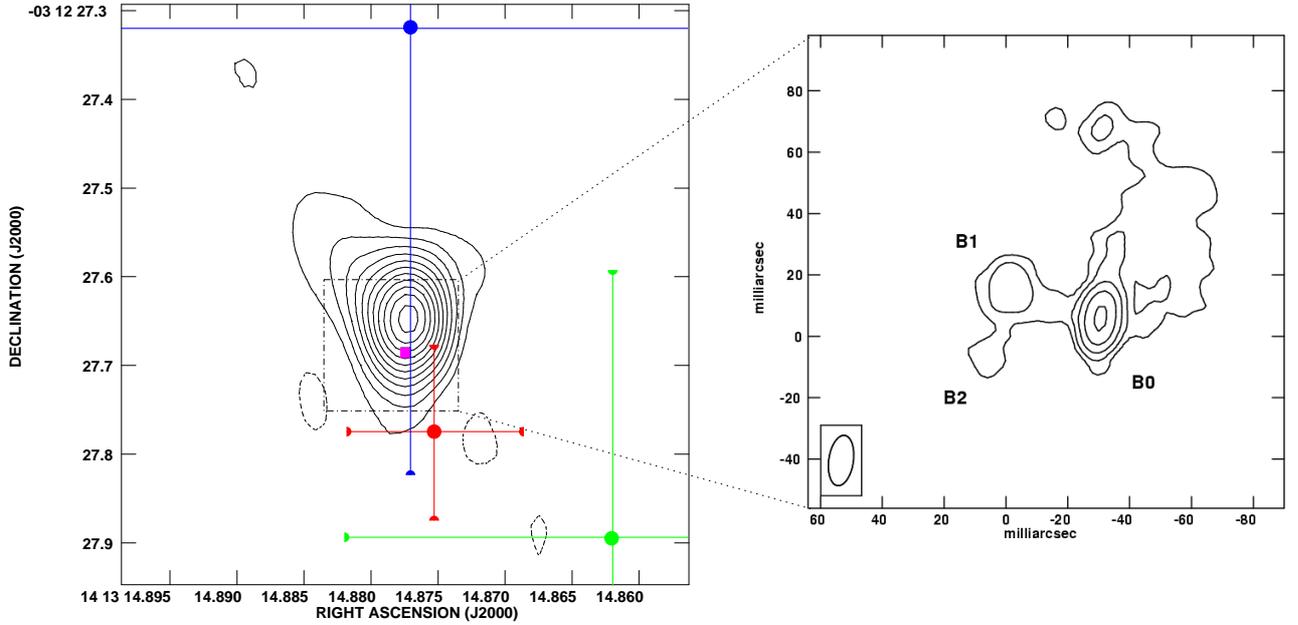}
\caption{{\it Left panel}: VLA A-array K-band continuum image of the nuclear region of NGC~5506. Contour levels are -1, 1, 2, 3, 4, 5, 6, 7, 8, 9, 10, 11\,$\times$\,2.1\,mJy\,beam$^{-1}$ (1$\sigma$ rms = 0.7\,mJy\,beam$^{-1}$). The beam size is 0\farcs12 $\times$ 0\farcs08. We show the positions of the optical center (blue circle, taken from \citealt{adelman08}), water maser emission in our VLA DnC-array (green circle) and VLA A-array (red circle) maps, respectively. The size of the bars indicates the uncertainty in the positions derived. The dotted-dashed square indicates the area covered by the VLBI map of \citet{middelberg04} and the magenta square the position of the brightest source, B0, in this map. {\it Right panel}: the EVN and MERLIN 6 cm image of the nuclear region of NGC~5506 by \citet{middelberg04}. Offsets are relative to RA$_{\rm J2000}$ = 14$^{\rm h}$ 13$^{\rm m}$ 14\fs87953, Dec$_{\rm J2000}$ = --03$^{\rm \circ}$ 12$^{\rm \prime}$ 27\farcs6799. The size of the beam, shown in the bottom-left corner, is 16.5 $\times$ 8.0 mas.}
\label{fig:trybig}
\end{figure*}

\subsubsection{NGC~5506:}
Figure~\ref{fig:trybig} shows the innermost region of NGC~5506 as seen from our VLA K-band image (left panel) and from the EVN+MERLIN\footnote{Multi-Element Radio Linked Interferometer Network} C-band map of \citet{middelberg04} (right panel). Overlaid on the K-band contour map are the positions of the main centres of activity in the galaxy, namely the optical nucleus and the brightest radio knot detected by  \citet{middelberg04}, in addition to the location of the maser emission derived in our VLA DnC and A-array maps. Within the uncertainties, all positions are coincident. \citet{middelberg04} proposed two main scenarios to describe their VLBI map: i) the strongest knot, B0, is a candidate location of the AGN core owing to its flat spectrum, compactness, and high brightness temperature ($\sim$ 3.6 $\times$ 10$^{8}$ K); ii) the AGN might lie between B0 and B1, hence these two bright components are the ``working surfaces'' where double-sided jets from the AGN impact the interstellar medium (ISM). The region of emission north of B0 and component B2 could be low-surface-brightness emission seen extending from the hot spots to a larger scale. The flat radio spectrum of B0 is indicative of optically thick synchrotron emission, while ``working surfaces'', as in compact symmetric objects (CSOs), have steeper spectra, which are typical of optically thin synchrotron emission, therefore favouring B0 as being the core. In any case, our spectral line observations indicate a position for the maser that is consistent with either one of the nuclear components, hence an association of the maser emission with an accretion disk, a radio jet, or a nuclear outflow. Indications of a one-sided jet and an outflow are found for this galaxy (\citealt{xant10}, and references therein) and the blue-shifted velocities of the maser lines w.r.t. the systemic velocity may be more consistent with a jet or an outflow origin. 

\citet{guainazzi10} measured the inclination of the nuclear disk in NGC5506 to be $i$=40\degr using the profile of the X-ray FeK$\alpha$ emission line. This measurement seems also consistent with the value of the nuclear column density (N$_{\rm H}$ $\sim$ 3$\pm$1\,$\times$\,$10^{22}$ cm$^{-2}$) for this galaxy obtained by \citet{bianchi03}, which is in-between those typically found for Sy\,1 and Sy\,2 galaxies (N$_{\rm H}$ $\le$ $10^{22}$ and N$_{\rm H}$ $\sim$ $10^{23}$ cm$^{-2}$, respectively; e.g. \citealt{malizia09}). Such an inclination for the disk in NGC~5506 would be less favourable for producing detectable (disk) maser emission (see also Sect.~\ref{detection}). Furthermore, if we also assume that such an inclination can still provide a long enough amplification path for maser action, we would obtain a probability (P($i$)=cos($i$)) of detecting water masers in AGN accretion disks of $\sim$ 77\%\footnote{This estimate is based on simple geometrical considerations, thus assumes that all AGN accretion disks host maser emission and ignores the many physical parameters on which the detection of masers depends.}. This value is much higher than any detection rate obtained in AGN surveys. Under the same assumptions made before and in the (unlikely) case that in NLS1s we only have disk-masers, we can then derive the range of accretion disks inclinations allowed to account for the maser detection rate found in NLS1s of $\sim$ 7\% (or $\sim$ 20\% when considering only targets within 10000 \kms; see Sect.~\ref{detection}). With these input parameters, the inclination of the nuclear disk favourable to producing maser action in NLS1s should not be less than 86\degr (or 79\%). All this seems to support the conclusion that the maser in NGC~5506 is not associated with an accretion disk, favouring instead an association with the radio jet or a nuclear outflow, or an atypical maser case.

\begin{figure}
\centering
\includegraphics[width= 9 cm]{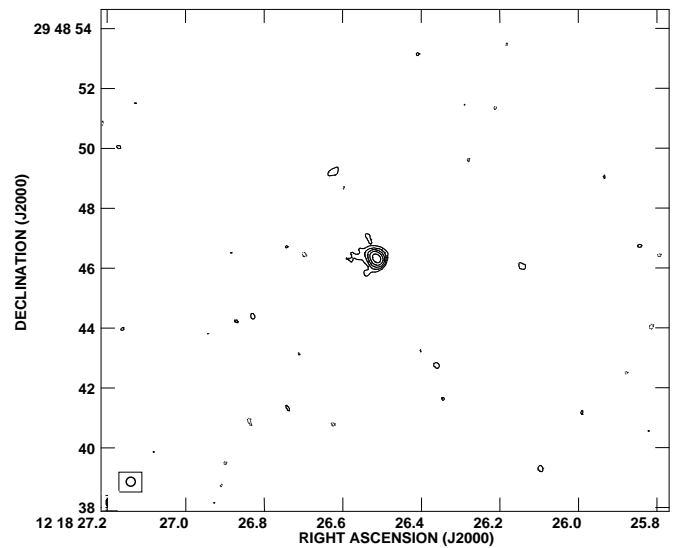}
\caption{VLA A-array X-band naturally-weighted continuum image of Mrk~766. Contour levels are -1, 1, 2, 4, 8, 16\,$\times$0.2\,mJy\,beam$^{-1}$ (1$\sigma$ rms = 0.07\,mJy\,beam$^{-1}$). The beam size, shown in the lower-left corner is 0\farcs28 $\times$ 0\farcs24.}
\label{n4253}
\end{figure}

\subsubsection{Mrk~766}
For Mrk~766, neither interferometric measurements of the maser emission nor a VLBI map of the nuclear radio continuum are available. From VLA A-array archival data, we have produced a radio continuum map at X-band of the nucleus of Mrk~766. The image shows a compact nuclear source of $\sim$ 5 mJy (Fig.~\ref{n4253}). We have been recently granted EVLA A-array time to observe the maser emission in Mrk~766. These measurements will allow us to compare the position of the maser with that of the nuclear radio continuum source(s), and thereby discriminate between a nuclear maser and a (less likely) maser associated with a star-forming region in the large-scale disk of the galaxy. We note an analysis of X-ray Multi-mirror Mission (XMM) time-resolved X-ray spectroscopic data, has found in Mrk~766 evidence of gas in a Keplerian orbit around a (super)massive black hole \citep{turner06}.

\subsubsection{IGR~J16385}
For IGR~J16385, only the 1.4 GHz radio continuum map from the National Radio Astronomy Observatory VLA Sky Survey (NVSS) at 1.4 GHz is available. The NVSS map shows a radio source of $\sim$ 10 mJy at the position of the optical peak of the galaxy. Expanded VLA (EVLA) B-array observations of the radio continuum (X-Band) and maser line (K-Band) in IGR~J16385 have been requested and await evaluation.

\subsubsection{IRAS~0345}
This galaxy has not been well studied. The nuclear radio continuum has been imaged by \citet{thean00} at 8.4 GHz with the VLA A-array, showing a single, slightly resolved source with an integrated flux of 6.8 mJy (Thean et al. 2000; their Fig.~1). The galaxy was also part of the sample studied by \citet{thompson09} using Spitzer mid-infrared (MIR) spectroscopic data and their conclusion was that its observational characteristics are consistent with a nucleus obscured by a compact and clumpy toroidal structure of dust. Owing to the lack of radio interferometric measurements, however, the association of the maser emission with the AGN or the nuclear starburst present in this galaxy cannot yet be determined. Indeed, IR spectroscopic data of a sample of Seyfert galaxies that includes IRAS~0345 were analysed by \citet{imanishi04}. Their work indicates that nuclear starbursts occurring in the dusty tori of Seyfert galaxies may be physically connected to the central AGN. This view is quite different from that of the classical unification paradigm, in which the dusty tori 'simply' hide the central AGN of Sy\,2 galaxies and reprocess AGN radiation as infrared dust emission in Seyfert galaxies.\\

While water masers are not typically detected in Sy\,1 galaxies, our work demonstrates that they {\it are} detected in NLS1 galaxies, at a rate comparable to Sy\,2's.  Thus it is clear that the NLS1 class and the maser phenomenon are somehow associated.  The details of this association are unclear.  The next step to investigating the nature of masers in NLS1 galaxies would be to image the masers on parsec or sub-parsec scales with VLBI observations.


The low black hole masses measured in NLS1 galaxies have suggested that NLS1s have accretion rates close to the maximum allowed values, causing their observed extreme behaviour (\citealt{marconi08}, and references therein). Owing to their low black hole masses, these objects may represent an important link between AGN and the elusive intermediate-mass black holes \citep{komossa08}. Strong radiation-pressure outflows represents another possible explanation of the peculiar characteristics of NLS1 galaxies. In particular, estimates of black hole masses turn out to be severely underestimated when the effect of radiation pressure is neglected, thus making NLS1s more extreme than they actually are (e.g., \citealt{marconi08}). Another possibility that may account for the NLS1 properties is that they may be viewed at angles intermediate between type 1 and type 2 Seyfert galaxies. Nuclear obscuration may thus play an important role in determining the observed properties of NLS1s.

Compared to the general AGN population, the five galaxies examined in our study have low black hole masses and high accretion rates\footnote{These estimates, however, need to be interpreted with caution since they strongly depend on the method used and the precision to which the distance of the galaxy is known.}, consistent with their classification as NLSy1 galaxies (Table~\ref{lines}). A direct association between the high detection rate of water masers in NLS1 and the two aforementioned peculiar quantities in this class of AGN cannot be presently assessed and requires a more complete approach. At this stage, it seems instead that there is a possible relation between the maser action and the presence of nuclear outflows detected in NLS1. As mentioned in Sect.~\ref{origin}, the water maser in Circinus, which has similar spectral properties to some maser spectra in our NLS1s, is associated with an outflow produced by the AGN activity (\citealt{greenhill03}). The association of the maser emission in NLS1 with favourably orientated outflows may be corroborated by the blue-shifted velocities of the main maser features w.r.t. the systemic velocity of the galaxy in four out of five of our targets (see Table~\ref{lines} and Sect.~\ref{origin}). As shown in Sect.~\ref{origin} for the case of NGC~5506, it is instead more controversial, although not implausible, that the intermediate inclination of nuclear disks and tori proposed for NLS1s can be related to the production of water masers. However, the presence of nuclear optically thick gas and/or the partial covering of the nucleus, possibly related to the aforementioned disks/tori inclination, has been invoked to describe the X-ray characteristics of a number of NLS1s, including our target galaxies (\citealt{guainazzi10} for NGC~5506, \citealt{turner06} for Mrk~766, and \citealt{terashima09} for NGC~4051). Given the relation between nuclear obscuration and maser emission reported for relatively large samples of maser galaxies (hence, mostly type 2 Seyfert) by, e.g., \citet{greenhill08}, \citet{zhang06a}, and \cite{zhang10}, this could partly justify the relatively high maser detection rate in NLS1s, especially w.r.t. that in Sy\,1.

\section{Summary}

We have observed a sample of NLS1 galaxies with the GBT in a search for water maser emission at 22 GHz. We have also analysed GBT and VLA data and produced 22-GHz spectra for the five NLS1 galaxies hosting maser emission. 

The main results are: i) an overall water maser detection rate in NLS1 galaxies of $\sim$ 7\% for the full sample and a detection rate of $\sim$ 21\% for a volume-limited (V$<$10000 km/s) sample, demonstrating a relation between NLS1s and water maser phenomenon; ii) a strong indication that water maser emission in NLS1s is associated with their nuclear outflows or radio jet. 

Overall, this work shows the potential of water maser studies to unveil the innermost regions and investigate the main nuclear components of NLS1s in order to more clearly understand their peculiar characteristics, in particular, in the framework of the unified model.

%

\begin{acknowledgements}
We are grateful to an anonymous referee for his/her comments on the manuscript. AT would like to thank Matteo Murgia for useful discussion. This research has made use of the NASA/IPAC Extragalactic Database (NED), which is operated by the Jet Propulsion Laboratory, Caltech, under contract with NASA. This research has also made use of the NASA's Astrophysics Data System Abstract Service (ADS).
\end{acknowledgements}

\bibliographystyle{aa} 
\bibliography{17213pap} 

\onllongtab{1}{
\begin{longtable}{lcccccccc}
 \caption{\label{all} NLS1 galaxies for which there have been searches for water maser emission. Standard text font: our new sample taken from \citet{veron01}; in italics: sources observed in previous surveys; bold face: objects displaying water maser emission.}\\
\hline\hline
Source$^{a}$& R.A.$^{b}$ & Dec.$^{b}$ &$V_{sys}^{c}$& Obs.\ Date$^{d}$ & T$_{\rm int}$ &  $\Delta\,{\rm V}$ & rms$^{e}$   & $L_{\rm{H_{2}O}}^{f}$ \\ 
            &            &            &           & &            &                       &            &                 \\
            & (J2000)    & (J2000)    &   (km/s)  &  & (min)      & (km/s)                & (mJy)      & (\solum)        \\
\hline
\endfirsthead
\caption{continued.}\\
\hline\hline
Source$^{a}$& R.A.$^{b}$ & Dec.$^{b}$ &$V_{sys}^{c}$ & Obs.\ Date$^{d}$ & T$_{\rm int}$ &  $\Delta\,{\rm V}$ & rms$^{e}$    & $L_{\rm{H_{2}O}}^{e}$ \\ 
           &            &            &        &    &             &                       &            &                 \\
           & (J2000)    & (J2000)    &   (km/s)  & & (min)       & (km/s)                 & (mJy)     & (\solum)        \\
\hline
\endhead
\hline
\endfoot
IZw1                       &    00:53:34.9  &  +12:41:36  & 17658  & 2011-02-06  & 20 &     16153$-$19177	 &   2.3 &  $<$ 8.8   \\
TonS180                    &    00:57:20.2  &  -22:22:56  & 18581  & 2011-02-10  &20 &     17068$-$20109	 &   2.0 &  $<$ 8.5    \\
Mrk359                     &    01:27:32.5  &  +19:10:44  &  5212  & 2011-02-06  &20 &      3823$-$6614	 &   2.1 &  $<$ 0.7    \\
MS01442-0055               &    01:46:44.7  &  -00:40:43  & 24820  & 2011-02-10  &20 &     23247$-$26408	 &   1.8 &  $<$13.6    \\
Mrk1044                    &    02:30:05.5  &  -08:59:53  &  4932  & 2011-02-10  &20 &      3545$-$6331	 &   2.1 &  $<$ 0.6    \\
HS0328+0528                &    03:30:52.2  &  +05:38:26  & 13790  & 2011-02-12  &20 &     12322$-$15272	 &   1.9 &  $<$ 4.4    \\
{\bf {IRAS 03450+0055}}    &    03:47:40.1  &  +01:05:14  &  9294  & 2011-02-12  &20 &      7868$-$10734	 &   2.2 &  67  \\
IRAS04312+4008             &    04:34:40.9  &  +40:14:19  &  6141  & 2011-02-12  &20 &      4743$-$7551	 &   2.1 &  $<$ 1.0    \\
{\it {MRK618}}             &    04:36:22.2  &  -10:22:34  & 10658  & 2009-11-29  &10 &      9219$-$12110      &   3.7 &  $<$ 5.2    \\
IRAS04416+1215             &    04:44:28.8  &  +12:21:11  & 26652  & 2011-02-12  &20 &     25061$-$28258	 &   2.2 &  $<$19.2    \\
{\it{IRAS 04576+0912}}     &    05:00:20.8  &  +09:16:56  & 10822  & 2009-11-16  &10   &      9381$-$12276    &   4.6 &  $<$ 6.6    \\
IRAS04596-2257             &    05:01:48.7  &  -22:53:24  & 12232  & 2011-02-12  &20 &     10778$-$13699	 &   2.6 &  $<$ 4.8    \\
{\it{IRAS 05262+4432}}     &    05:29:55.5  &  +44:34:39  &  9644  & 2009-12-30  &10   &      8214$-$11087    &   2.5 &  $<$ 2.8    \\
{\it{UGC  3478}}           &    06:32:47.2  &  +63:40:25  &  3830  & 2005-12-22  & 5  &      2453$-$5219      &   5.3 &  $<$ 0.9    \\
RXJ07527+2617              &    07:52:45.5  &  +26:17:37  & 24632  & 2011-02-09  &20 &     23061$-$26218	 &   1.9 &  $<$14.1    \\
Mrk382                     &    07:55:25.3  &  +39:11:10  & 10099  & 2011-02-09  &20 &      8665$-$11546	 &   1.9 &  $<$ 2.4    \\
 {\it{MCG +04.22.042}}     &    09:23:43.0  &  +22:54:32  &  9705  & 2008-12-14  &40   &      8268$-$11141    &   1.7 &  $<$ 2.0    \\
 {\it{MARK  110}}          &    09:25:12.9  &  +52:17:11  & 10580  & 2008-02-08  &10   &      9142$-$12032    &   3.8 &  $<$ 5.2    \\
Mrk705                     &    09:26:03.3  &  +12:44:03  &  8739  & 2011-02-09  &20 &      7318$-$10173	 &   2.1 &  $<$ 2.0    \\
Mrk707                     &    09:37:01.1  &  +01:05:43  & 15091  & 2011-02-09  &20 &     13611$-$16585	 &   2.0 &  $<$ 5.6    \\
 {\it{Zw 063.060}}         &    09:45:29.3  &  +09:36:10  &  3988  & 2009-12-18  &10   &      2610$-$5378     &   4.0 &  $<$ 0.8    \\
Mrk124                     &    09:48:42.6  &  +50:29:31  & 16878  & 2011-02-09  &20 &     15381$-$18389	 &   1.9 &  $<$ 6.6    \\
Mrk1239                    &    09:52:19.1  &  -01:36:44  &  5974  & 2011-02-09  &20 &      4578$-$7383	 &   2.2 &  $<$ 1.0    \\
IRAS09571+8435             &    10:05:17.0  &  +84:20:44  & 27641  & 2011-02-12  &20 &     26040$-$29257	 &   2.0 &  $<$18.7    \\
PG1011-040                 &    10:14:20.7  &  -04:18:39  & 17482  & 2011-02-09  &20 &     15979$-$18999	 &   1.9 &  $<$ 7.1    \\
PG1016+336                 &    10:19:49.5  &  +33:22:04  &  7345  & 2011-02-09  &20 &      5936$-$8766	 &   2.0 &  $<$ 1.3    \\
 {\it{MARK  142}}          &    10:25:31.3  &  +51:40:35  & 13474  & 2010-02-01  &10   &     12009$-$14953    &   2.6 &  $<$ 5.8    \\
KUG1031+398                &    10:34:38.6  &  +39:38:29  & 12724  & 2011-02-09  &20 &     11266$-$14196	 &   1.9 &  $<$ 3.8    \\
RXJ10407+3300              &    10:40:43.9  &  +33:00:59  & 24512  & 2011-02-09  &20 &     22942$-$26097	 &   2.0 &  $<$14.7    \\
Mrk734                     &    11:21:47.1  &  +11:44:19  & 15050  & 2011-02-09  &20 &     13570$-$16544	 &   2.2 &  $<$ 6.1    \\
Mrk739E$^{*}$              &   11:36:29.3  &   +21:35:45  &  8909  &  2011-02-09  &20 &      7486$-$10345	 &   1.9 &  $<$ 1.8    \\
 {\it{MCG +06.26.012}}     &    11:39 13.9  &  +33:55:51  &  9819  & 2010-02-01  &10   &      8388$-$11263    &   2.4 &  $<$ 2.8    \\
Mrk42                      &    11:53:41.8  &  +46:12:43  &  7385  & 2011-02-10  &20 &      5976$-$8807	 &   2.2 &  $<$ 1.5    \\
\textbf{\emph{NGC 4051}}   &    12:03:09.6  &  +44:31:53  &   700  & 2004-01-31  &20 &     -647$-$2060        &   2.5  &   2  \\
PG1211+143                 &    12:14:17.7  &  +14:03:13  & 24253  & 2011-02-09  &20 &     22685$-$25835	 &   2.0 &  $<$14.4    \\
\textbf{\emph{MARK  766}}  &    12:18 26.5  &  +29:48:47  &  3876  & 2008-03-02  &60 &    2500$-$5266         &  1.5     &   8  \\
MS12170+0700               &    12:19:30.9  &  +06:43:34  & 24157  & 2011-02-09  &20 &     22590$-$25739	 &   1.9 &  $<$13.6    \\
MS12235+2522               &    12:26:04.2  &  +25:06:38  & 20086  & 2011-02-10  &20 &     18558$-$21628	 &   2.1 &  $<$10.4    \\
IC3599                     &    12:37:41.2  &  +26:42:28  &  6459  & 2011-02-10  &20 &      5059$-$ 7872	 &   2.0 &  $<$ 1.0    \\
PG1244+026                 &    12:46:35.3  &  +02:22:08  & 14443  & 2011-02-09  &20 &     12969$-$15931	 &   2.0 &  $<$ 5.1    \\
 {\it{NGC 4748}}           &    12:52:12.4  &  -13:24:53  &  4386  & 2009-12-16  &10    &     3004$-$5780     &   3.5 &  $<$ 0.8    \\
Mrk783                     &    13:02:58.8  &  +16:24:27  & 20146  & 2011-02-09  &20 &     18618$-$21688	 &   1.9 &  $<$ 9.5    \\
R14.01                     &    13:41:12.9  &  -14:38:40  & 12528  & 2011-02-09  &20 &     11072$-$13998	 &   2.2 &  $<$ 4.2    \\
Mrk69                      &    13:46:08.1  &  +29:38:10  & 22969  & 2011-02-09  &20 &     21414$-$24539	 &   1.9 &  $<$12.3    \\
2E1346+2646                &    13:48:34.9  &  +26:31:10  & 17660  & 2011-02-09  &20 &     16156$-$19179      &   2.1 &  $<$ 8.0    \\
PG1404+226                 &    14:06:21.9  &  +22:23:46  & 29380  & 2011-02-09  &20 &     27763$-$31013	 &   2.4 &  $<$25.4    \\
\textbf{\emph{NGC5506}}    &    14:13:14.8  &  -03:12:26  &  1853  & 2008-10-27  &15 &     494$-$3224     &   3.1    &   24  \\
 {\it{MARK  684}}          &    14:31:04.8  &  +28:17:14  & 13814  & 2010-02-01  &10   &     12346$-$15296    &   2.7 &  $<$ 6.3    \\
Mrk478                     &    14:42:07.5  &  +35:26:23  & 23700  & 2011-02-09  &20 &     22138$-$25277	 &   2.1 &  $<$14.5    \\
PG1448+273                 &    14:51:08.8  &  +27:09:27  & 19487  & 2011-02-09  &20 &     17965$-$21023	 &   2.1 &  $<$ 9.8    \\
IRAS15091-2107             &    15:11:59.8  &  -21:19:02  & 13373  & 2011-02-09  &20 &     11909$-$14851	 &   2.4 &  $<$ 5.3    \\
MS15198-0633               &    15:22:28.8  &  -06:44:41  & 24919  & 2011-02-09  &40 &     23345$-$26508	 &   1.5 &  $<$11.4    \\
Mrk486                     &    15:36:38.3  &  +54:33:33  & 11672  & 2011-02-10  &30 &     10224$-$13134	 &   1.5 &  $<$ 2.5    \\
IRAS15462-0450             &    15:48:56.8  &  -04:59:34  & 29917  & 2011-02-09  &20 &     28294$-$31555	 &   2.5 &  $<$27.4    \\
 {\it{MARK  493}}          &    15:59:09.6  &  +35:01:47  &  9392  & 2010-01-29  &10   &      7965$-$10832    &   2.6 &  $<$ 2.8    \\
{\bf {IGR J16385-2057}}    &    16:38:30.9  &  -20:55:24  &  8064  & 2010-03-28  &40 &  6649$-$9492          &   2.0    &  10  \\ 
EXO16524+3930              &    16:54:08.1  &  +39:25:33  & 20656  & 2011-02-09  &15 &     19123$-$22203	 &   2.3 &  $<$12.0    \\
B31702+457                 &    17:03:30.3  &  +45:40:48  & 18107  & 2011-02-10  &20 &     16598$-$19630	 &   1.8 &  $<$ 7.2    \\
RXJ17450+4802              &    17:45:04.5  &  +48:02:40  & 16189  & 2011-02-10  &10 &     14698$-$17693	 &   2.8 &  $<$ 9.0    \\
Kaz163                     &    17:46:59.1  &  +68:36:28  & 18887  & 2011-02-12  &15 &     17371$-$20417	 &   2.9 &  $<$12.7    \\
Mrk507                     &    17:48:38.4  &  +68:42:16  & 16758  & 2011-02-12  &15 &     15262$-$18268	 &   3.0 &  $<$10.3    \\
HS1817+5342                &    18:18:10.4  &  +53:43:46  & 23983  & 2011-02-12  &10 &     22418$-$25563	 &   4.4 &  $<$31.0    \\
HS1831+5338                &    18:32:49.7  &  +53:40:22  & 11692  & 2011-02-12  &10 &     10243$-$13154      &   5.0 &  $<$ 8.4    \\
Mrk896                     &    20:46:20.8  &  -02:48:45  &  7922  & 2011-02-06  &20 &      6508$-$ 9349	 &   3.4 &  $<$ 2.6    \\
 {\it{IGR J21277+5656}}    &    21:27:45.4  &  +56:56:33  & 4317   & 2008-01-27  &15   &      2936$-$5710     &   3.0 &  $<$ 0.7    \\
MS22102+1827               &    22:12:37.0  &  +18:42:28  & 24498  & 2011-02-10  &20 &     22928$-$26083	 &   1.9 &  $<$14.0    \\
Akn564                     &    22:42:39.3  &  +29:43:32  &  7400  & 2011-02-10  &20 &      5991$-$8822	 &   1.9 &  $<$ 1.3    \\
HS2247+1044                &    22:49:39.5  &  +11:00:29  & 24883  & 2011-02-06  &20 &     23309$-$26472	 &   2.3 &  $<$17.5    \\
 {\it{MS 22549-3712}}      &    22:57:38.9  &  -36:56:07  & 11692  & 2010-01-14  &10   &     10243$-$13154    &   5.9 &  $<$ 9.9    \\
Kaz320$^{*}$               &     22:59:33.0  &  +24:55:06  & 10343 &  2011-02-06  &20 &      8907$-$11792	 &   2.6 &  $<$ 3.4    \\
 {\it{SDSS J23386-0028}}   &    23:38:37.1  &  -00:28:10  & 10792  & 2006-02-07  &10   &      9352$-$12246    &   3.2 &  $<$ 4.6    \\

\hline
\end{longtable}
\noindent$^a$ Source names and classification are taken from \citet{veron01} and \citet{veron01}.  Source names format: standard text = sources from this work; italics = sources from past surveys; boldface= sources from any survey where \water\ maser has been detected.\\
$^b$ Coordinates of the galaxies observed for this work are taken from Veron-Cetty \& Veron (2010). For sources targeted by searches for water maser emission in past surveys, we used the coordinates of Braatz et al.\ (in prep).\\
$^c$ LSR velocities of the galaxies observed for this work are taken from the NASA/IPAC Extragalactic Database (NED). For sources in past surveys we used LSR velocities from Braatz et al.\ (in prep).\\
$^{d}$ All observations were made with the GBT.\\  
$^{e}$ This is the 1$\sigma$ rms of the spectrum measured for a 24 kHz (0.33 \kms) channel, after Hanning smoothing.\\
$^{f}$ Water maser luminosity thresholds have been calculated from the 3$\sigma$ rms level by assuming a 1 \kms\ line width and a value for H$_{0}$=75 \kms\/Mpc. Luminosities for the detected maser sources refer to the total integrated maser luminosity, as reported in Table~\ref{lines}.\\
$^{*}$ These galaxies were re-observed by ourselves, although they had previously been targeted by searches for water masers at slightly different coordinates and velocities w.r.t. ours (Mrk739E: RA(J2000)=11:36:29.1; Dec.(J2000)=+21:35:46; $V_{sys}$=8956 \kms. Kaz320: RA(J2000)=22:59:32.9; Dec.(J2000)=+24:55:06; $V_{sys}$=10350 \kms).}

\begin{table*}
\centering
\caption{Single-dish water maser line characteristics. Line parameters have been derived for individual components through Gaussian fit procedures implemented in GBTIDL.}
\label{lines}
\begin{minipage}[t]{\textwidth}
\renewcommand{\footnoterule}{}

\begin{tabular}{lcccccccccc}
\hline\hline

Source    & $cz$$^{a}$     & M$_{BH}$$^{b,*}$ & L$_{\rm bol}$/L$_{\rm Edd}$$^{c,*}$ & L$_{\rm X}$$^{d,*}$ & S$^{peak}_{H_{2}O}$ & V$^{peak}_{H_{2}O}$ & ${\Delta}$v & S$^{int}_{H_{2}O}$  & L$^{iso}_{H_{2}O}$ & $V_{H_{2}O}^{Hel}-V_{sys}^{Hel}$ \\
          & (km/s)   &     &           &         & (mJy/b)    &   (\kms)      &   (\kms)   &   (Jy/b*\kms)  &   (\solum)             &  (\kms)   \\

\hline 
NGC\,4051 & 700     & 0.5$^{+6}_{-3}$      & 0.57   & 0.15 & 13$\pm$2        & 651.8$\pm$0.2   & 2.7$\pm$0.4   & 0.04$\pm$0.01   & 0.08$\pm$0.02   &  $-$48 \\
           &          &            &        &      & 22$\pm$7        & 664.8$\pm$0.1   & 2.6$\pm$0.5   & 0.06$\pm$0.03   &  0.11$\pm$0.05  & $-$35 \\
            &         &            &        &      & 20$\pm$1        & 668.5$\pm$0.5   & 6$\pm$2       & 0.12$\pm$0.05   &  0.22$\pm$0.09  & $-$31 \\
           &          &            &        &      & 48$\pm$4        & 672.39$\pm$0.03 & 1.6$\pm$0.1   & 0.08$\pm$0.01   &  0.14$\pm$0.02  & $-$28 \\
            &         &            &        &      & 24$\pm$2        & 675.03$\pm$0.06 & 1.5$\pm$0.1   & 0.036$\pm$0.005 & 0.07$\pm$0.01 & $-$25 \\
           &          &            &        &      & 20$\pm$3        & 693.6$\pm$0.1   & 1.5$\pm$0.3   & 0.03$\pm$0.01   & 0.06$\pm$0.02   & $-$6 \\
          &           &            &        &      & 17$\pm$2        & 703.0$\pm$0.2   & 4.2$\pm$0.6   & 0.07$\pm$0.02   & 0.13$\pm$0.03   & $+$3 \\
           &          &            &        &      & 63$\pm$4        & 711.77$\pm$0.03 & 0.93$\pm$0.07 & 0.059$\pm$0.008 & 0.11$\pm$0.02 & $+$12 \\
          &           &            &        &      & 19.2$\pm$0.8    & 721.1$\pm$0.7   & 26$\pm$2      & 0.50$\pm$0.06   & 0.9$\pm$0.1     & $+$21 \\
          &           &            &        &      & 82$\pm$3        & 732.33$\pm$0.03 & 1.61$\pm$0.07 & 0.13$\pm$0.01   & 0.25$\pm$0.02   & $+$32 \\
          &           &            &        &      & 20$\pm$2        & 765.12$\pm$0.07 & 1.3$\pm$0.2   & 0.026$\pm$0.007 & 0.05$\pm$0.01 & $+$65 \\
            &       &            &        &      &                 &                 &               &                 & (sum =) {\bf{2}} &      \\
Mrk~766 & 3876    & 3          & 0.65   & {\bf {7}}    & 11.2$\pm$0.4    & 3792.2$\pm$0.2  &  11.5$\pm$0.5 &  0.13$\pm$0.01  & 8.0$\pm$0.6 & $-$84 \\
            &       &            &        &      &                 &                 &               &                 & (sum =) {\bf{8}} &      \\
NGC\,5506 & 1853    & 5$^{+2}_{-1}$          & 0.18   & 6    & 11$\pm$1       & 1724.7$\pm$0.2   & 5.0$\pm$0.6  &  0.06$\pm$0.01 &  0.8$\pm$0.1   & $-$128 \\
          &         &            &        &      & 18$\pm$3        & 1786.07$\pm$0.08& 1.0$\pm$0.2   & 0.018$\pm$0.007 & 0.3$\pm$0.1   & $-$67 \\ 
          &         &            &        &      & 20$\pm$3        & 1798.95$\pm$0.08& 1.0$\pm$0.2   & 0.020$\pm$0.007 & 0.3$\pm$0.1   & $-$54 \\
          &         &            &        &      & 12.5$\pm$0.3    & 1805$\pm$1      &  112$\pm$3    &   1.4$\pm$0.7   & 20$\pm$10    & $-$48 \\
          &         &            &        &      & 21$\pm$2        & 1822.7$\pm$0.1  & 2.0$\pm$0.3   & 0.04$\pm$0.01   &  0.6$\pm$0.1    & $-$30 \\
          &         &            &        &      & 27$\pm$3        & 1837.1$\pm$0.1  & 2.8$\pm$0.2   & 0.08$\pm$0.01   &  1.1$\pm$0.1    & $-$16 \\
          &         &            &        &      & 19$\pm$2        & 1848.4$\pm$0.2  & 3.4$\pm$0.4   & 0.06$\pm$0.01   &  0.8$\pm$0.1    & $-$5  \\
          &         &            &        &      & 12$\pm$2        & 1916.8$\pm$0.1   & 2.3$\pm$0.4  & 0.028$\pm$0.009 & 0.4$\pm$0.1   & $+$64  \\
            &       &            &        &      &                 &                 &               &                 & (sum =) {\bf{24}} &      \\
IGR\,J16385 & 8064  & 7          & 1.2    & 7    & 17$\pm$2        & 8035.29$\pm$0.06&  1.2$\pm$0.1  & 0.020$\pm$0.004 &  5$\pm$1    & $-$29 \\
            &       &            &        &      & 6$\pm$1         & 8057.4$\pm$0.1  &  1.0$\pm$0.3  & 0.006$\pm$0.003 & 1.6$\pm$8   & $-$7 \\
            &       &            &        &      & 9$\pm$1         & 8060.3$\pm$0.1  &  1.4$\pm$0.2  & 0.013$\pm$0.003 & 3.5$\pm$8   & $-$4 \\
            &       &            &        &      &                 &                 &               &                 & (sum =) {\bf{10}} &      \\
IRAS\,03450 & 9294  & 3          & 0.51   & 2.24$^{+1.3}_{-1.3}$ & 18$\pm$1        & 9257.7$\pm$0.3  &  10.3$\pm$0.7 & 0.19$\pm$0.02   &  67$\pm$7    & $-$36 \\
            &       &            &        &      &                 &                 &               &                 & (sum =) {\bf{67}} &      \\

\hline\\
\end{tabular}
\\
$^{a}$ Redshifts: NGC\,4051 from \citet{verheijen01}; Mrk~766 from \citet{smith87}; NGC\,5506 from \citet{keel96}; IGR\,J16385 from \citet{masetti08}; IRAS\,03450 from \citet{hewitt91}.\\  
$^{b}$ Black hole masses in units of 10$^{6}$ \solmass: NGC\,4051 from \citet{uttley04}; Mrk~766 from \citet{pounds03}; NGC\,5506 from \citet{nikolajuk09}; IGR\,J16385 from \citet{masetti08}; IRAS\,03450 from \citet{zhang06b}.\\
$^{c}$ Bolometric luminosity, L$_{\rm bol}$: NGC\,4051 and Mrk~766 from \citet{woo02}; NGC\,5506 from \citet{bianchi03}; IGR\,J16385 from \citet{malizia08}; IRAS\,03450 from \citet{zhang06b}.\\
$^{d}$ X-ray luminosity between 2-10 keV, L$_{\rm x}$, in units of 10$^{42}$ erg/s: NGC\,4051 from \citet{nucita10}; Mrk~766 from  \citet{pounds03}; NGC\,5506 from \citet{bianchi03}; IGR\,J16385 from \citet{panessa11};  for IRAS\,03450 the value for L$_{\rm x}$, taken from \citet{rush96}, refers to the 0.1-2.4 keV band.\\
$^{*}$ Errors in measured quantities, when available, are taken from the quoted paper. The total error, however, is dominated by the systematic uncertainties caused by the method used to estimate the reported value and/or the uncertainty in the distance to the target galaxy.
\end{minipage}
\end{table*}

\end{document}